\documentclass[fleqn,10pt]{wlscirep}
\usepackage[utf8]{inputenc}
\usepackage[T1]{fontenc}
\usepackage{braket}
\usepackage{amsmath}
\usepackage{multirow}
\usepackage{here}

\title{Conflict-free collective stochastic decision making by orbital angular momentum of photons through quantum interference}

\author[1,*]{Takashi Amakasu}
\author[1,2]{Nicolas Chauvet}
\author[3]{Guillaume Bachelier}
\author[3]{Serge Huant}
\author[1,2]{Ryoichi Horisaki}
\author[1,2,*]{Makoto Naruse}
\affil[1]{Department of Mathematical Engineering and Information Physics, Faculty of Engineering, The University of Tokyo, 7-3-1 Hongo, Bunkyo-ku, Tokyo 113-8656, Japan.}
\affil[2]{Department of Information Physics and Computing, Graduate School of Information Science and Technology, The University of Tokyo, 7-3-1 Hongo, Bunkyo-ku, Tokyo 113-8656, Japan.}
\affil[3]{Université Grenoble Alpes, CNRS, Institut Néel, 38042, Grenoble, France.}

\affil[*]{T.A. (email: amakasutakashi@gmail.com), M.N. (email: makoto\_naruse@ipc.i.u-tokyo.ac.jp)}

\begin{abstract}
In recent cross-disciplinary studies involving both optics and computing, single-photon-based decision-making has been demonstrated by utilizing the wave-particle duality of light to solve multi-armed bandit problems. Furthermore, entangled-photon-based decision-making has managed to solve a competitive multi-armed bandit problem in such a way that conflicts of decisions among players are avoided while ensuring equality. However, as these studies are based on the polarization of light, the number of available choices is limited to two, corresponding to two orthogonal polarization states. Here we propose a scalable principle to solve competitive decision-making situations by using the orbital angular momentum of photons based on its high dimensionality, which theoretically allows an unlimited number of arms. Moreover, by extending the Hong-Ou-Mandel effect to more than two states, we theoretically establish an experimental configuration able to generate multi-photon states with orbital angular momentum and conditions that provide conflict-free selections at every turn. We numerically examine total rewards regarding three-armed bandit problems, for which the proposed strategy accomplishes almost the theoretical maximum, which is greater than a conventional mixed strategy intending to realize Nash equilibrium. This is thanks to the quantum interference effect that achieves no-conflict selections, even in the exploring phase to find the best arms.
\end{abstract}
\begin{document}

\flushbottom
\maketitle

\thispagestyle{empty}

\section*{Introduction}
Optics and photonics are expected to play crucial roles in future computing systems \cite{kitayama2019novel}, making a variety of devices and systems to be intensively studied such as optical fibre-based neuromorphic computing \cite{de2019machine}, on-chip optical neural networks \cite{shen2017deep}, optical reservoir computing \cite{van2017advances}, among others. While these works are basically categorized in supervised learning, reinforcement learning is another important branch of artificial intelligence \cite{10.5555/551283}. The Multi-Armed Bandit (MAB) problem is an example of a reinforcement learning situation, which formulates a fundamental issue of decision making in dynamically changing uncertain environments where the target is to find the best selection among many slot machines, also referred to as arms, whose reward probabilities are unknown \cite{auer2002finite}. In solving MAB problems, exploration actions are necessary to find the best arm, although too much exploration may reduce the final amount of obtained reward from the exploitation. On the opposite, insufficient exploration may lead to miss the best arm. Furthermore, when multiple players are involved, decision conflicts become serious, as they induce congestions and inhibit socially achievable benefits \cite{lai2010cognitive,kim2016harnessing}. Equality among players is another critical issue, as unfair repartition of outcomes may lead to distrust the system. This whole problem is known as the competitive MAB (CMAB) problem. 

In order to solve these complex issues, photonic solutions have been recently considered. For example, the wave-particle duality of single photons has been utilized for the resolution of the two-armed bandit problem \cite{naruse2015single}. Moreover, Chauvet \textit{et al.} theoretically and experimentally demonstrated that polarization entangled photon pairs provide non-conflict and equality-assured decisions in two-player, two-armed bandit problems \cite{chauvet2019entangled}. Entangled photon states that allow more than three players while guaranteeing optimal outcome and equal repartition have also been demonstrated \cite{chauvet2020entangled}.

\begin{figure}[ht]
\centering
\includegraphics[width=12cm]{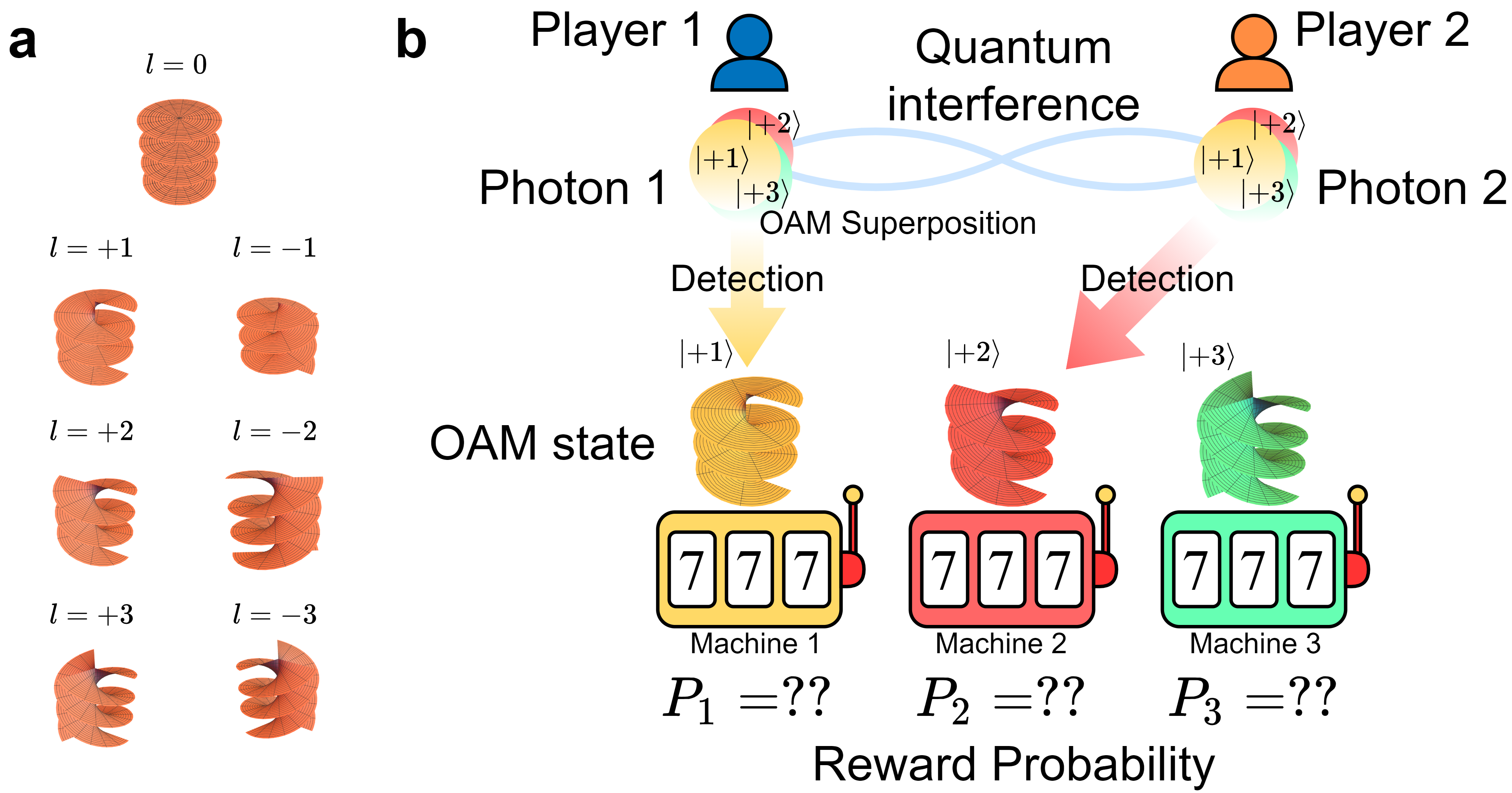}
\caption{\textbf{(a)} Schematic illustration of the phase surface of a beam with OAM. The direction of travel is upward. It is a surface with $|l|$ spirals. The pitch of each spiral is one wavelength divided by $|l|$. \textbf{(b)} Stochastic detection of OAM states directly corresponds to a player probabilistically selecting a machine.}
\label{fig:Concept}
\end{figure}

However, since these former principles rely on the polarization of light as the tunable degree of freedom, the number of possible selections or arms is limited to only two, although potential scalability for the single-player MAB is feasible within a tournament-based approach \cite{naruse2016single}. Therefore, the scalable principle of decision-making has been an important and fundamental issue, especially for multiplayer situations. 
In this paper, we introduce the use of the orbital angular momentum (OAM) of photons \cite{allen2003optical}, \cite{Forbes2021} to resolve the scalability issue of photonic decision making, following the concept summarized in Fig. \ref{fig:Concept}.

Photons that carry OAM \cite{allen2003optical} realize high-dimensional state spaces, only restricted by the precision and accuracy of the generation technique and the transmission medium \cite{yao2011orbital} (Fig. \ref{fig:Concept}a); hence one of the basic ideas of this study is to associate individual selections to different-valued OAM (Fig. \ref{fig:Concept}b). The applications of OAM have progressed in diverse areas ranging from the manipulation of cooled atoms, communications, nonlinear optics, optical solitons, and so on. The high-dimensionality of OAM is particularly attractive for quantum information processing in increasing the dimension of elementary quantum information carriers to go beyond the qubit \cite{Flamini2019}, \cite{Forbes2019}, \cite{Krenn2017}, \cite{Zhang2016}, \cite{Mirhosseini2015}, \cite{Molia-Terriza2007}. Likewise, in the present study, the multi-dimensionality of OAM plays a crucial role in extending the maximum number of arms as well as utilizing the probabilistic attribute of single photons carrying OAM.

Furthermore, to resolve CMAB problems when the number of arms is greater than two, we extend the notion of Hong-Ou-Mandel effect \cite{hong1987measurement} to more than two (OAM) vector states to induce quantum interference. We show that conflicting decisions among two players can be perfectly avoided by the adequate quantum interference design to generate OAM 2-photon states, relying on a coherent photon pair source. In the literature, OAM has been examined from game-theoretic perspectives such as resolving prisoners dilemma \cite{Pinheiro2013} and duel game \cite{Balthazar2015}. In the present study, we benefit from quantum interference for non-conflicting decision-making to maximize total rewards, which is similar to the insight gained by quantum game literature. Additionally, in solving CMAB problems with many arms, exploration action is necessary. We numerically examine total rewards regarding three-armed bandit problems where the proposed quantum-interference-based strategy accomplishes nearly theoretical maximum total reward. We confirm that the proposed strategy clearly outperforms conventional ones, including the mixed strategy intending to realize Nash equilibrium \cite{lai2010cognitive}.

Moreover, equality among players is important in CMAB problems. We demonstrate that equality is perfectly ensured by appropriate quantum interference constructions when the number of arms is three. At the same time, however, we also show that it is unfortunately impossible to accomplish perfect equality in the proposed scheme and with the current hypotheses when the number of arms is equal to or larger than four. Note also that perfect collision avoidance is ensured for any number of arms. 

These properties are made possible thanks to the high dimensionalities of OAM for scalability and the quantum interference effect for non-conflict selections even in the exploring phase to find the best arms.

\section*{Results}
\subsection*{Scalable decision maker with OAM}
\subsubsection*{System architecture for solving 1-player K-armed bandit problem.}
We first describe the problem under study, which is a stochastic multi-armed bandit problem with rewards following Bernoulli distributions defined as follows. There are $K$ available slot machines (or arms): when the player selects one arm $i$, the player wins with probability $P_i$ (and receives a fixed reward of 1), or loses with probability $1-P_{i}$ (and receives a fixed reward of 0), with $i$ an integer ranging from $1$ to $K$. Let a player choose an arm each time and allow a total of $T$ times, then the goal of the bandit problem is to find out which strategy should be followed to choose arms so that the resultant accumulated outcome is maximized. When the slot machine with the highest winning probability is known, the best strategy is to draw that specific arm for all $T$ times, but the player initially has no information about the arms. Therefore, exploration actions are required to know the best arm, whereas too much exploration potentially leads to missing a higher total amount of rewards from the best machine. 

In the previous work on single-photon decision maker using polarization \cite{naruse2015single}, two orthogonal linear polarizations of photons are associated with two slot machines; that is, horizontal and vertical polarizations correspond to slot machine 1 and 2, respectively. The exploration is physically realized by the probabilistic attribute of photon measurement, whose outcome depends on the direction of the polarization of linearly polarized single photons. Therein, the polarization degree of freedom physically and directly allows specifying the probabilistic selection of slot machines. However, as mentioned in the \textit{Introduction} section, the number of arms is limited to only two, although extendable in a single-player setup to powers of two via a tournament-based approach \cite{naruse2016single}.

\begin{figure}[ht] 
\centering
\includegraphics[width=\linewidth]{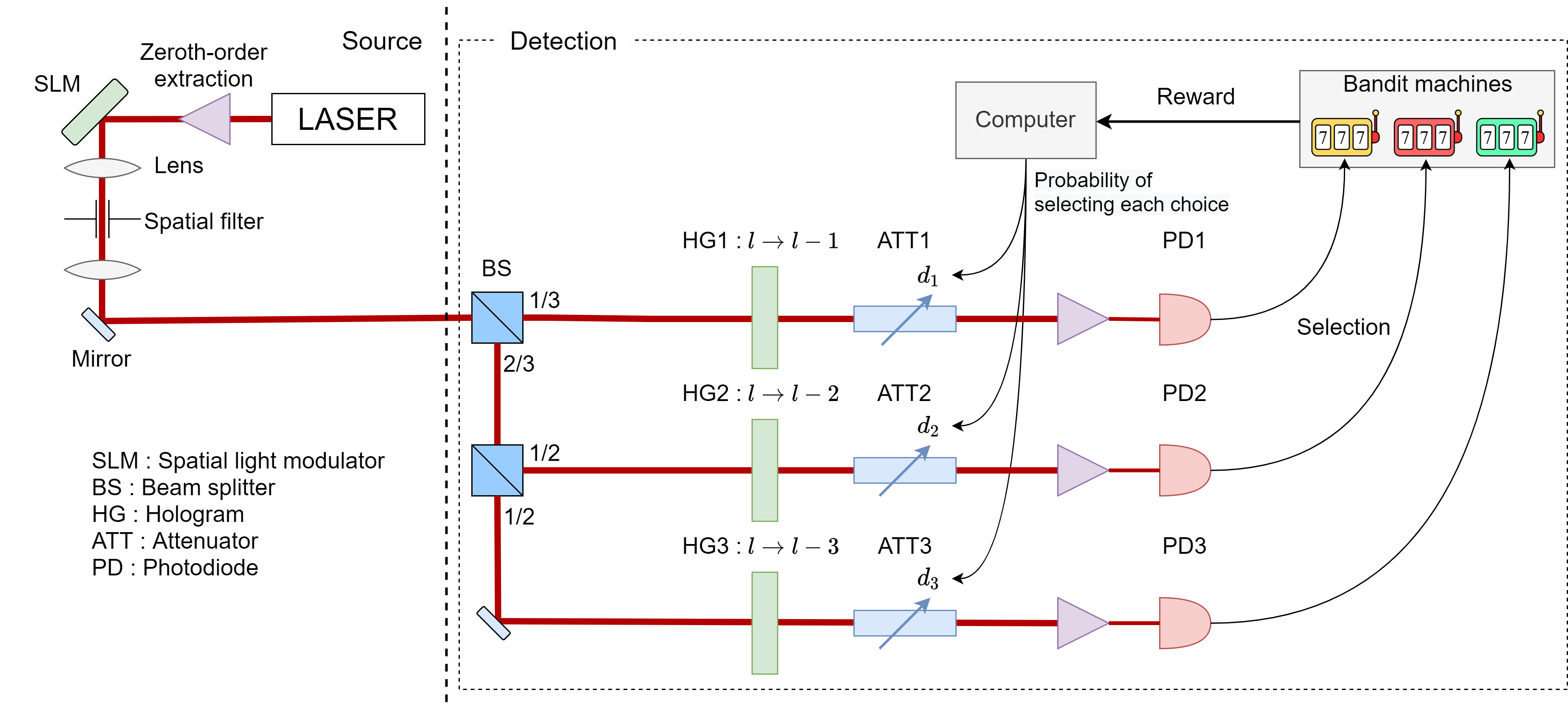}
\caption{Design of the system architecture for solving single-player 3-armed bandit problems. $d_1,d_2,d_3$ denote the transmittance of each attenuator.}
\label{fig:1System}
\end{figure}

The fundamental idea of the present study is to associate the dimension of OAM with the selection of multiple arms, whatever the number of arms. Allen \textit{et al.} have pointed out that a Laguerre-Gaussian (LG) beam has an angular momentum independent from polarization; they have called it OAM to distinguish it from the polarization-dependent spin angular momentum \cite{allen1992orbital}. The spatial mode of a LG beam can be expressed using the near-axis approximation
\begin{equation}
\begin{split}
u_l = f_m(\rho ,z)e^{i l\theta}e^{i k z} 
\end{split}
\end{equation}
where $\rho$ is the distance from the optical axis, $\theta$ is the azimuthal angle around the optical axis, $z$ is the coordinate of the propagation direction, $f_m$ is the complex amplitude distribution, and $k$ is the wavenumber. $m$ and $l$ are integer numbers that respectively describe the order of the Laguerre polynomial for the radial distribution and the azimuthal rotation number. In our study, $m$ is fixed at 0, while $l$ takes any integer numbers. Correspondingly, $\ket{l}$ is the state in which there is one photon in the $l$ mode, whose angular momentum is equal to $l \hbar$ where $\hbar$ is Planck's constant divided by $2\pi$. Since the modes with different $l$ are orthogonal to each other, the quantum state can be expressed by a linear superposition, using these modes as a basis. Figure \ref{fig:Concept}a schematically illustrates examples of beams with different $l$-valued OAM where $l$ is an integer from $-3$ to $3$. Non-zero $l$ beams exhibit spiral isophase spatial distributions. Figure \ref{fig:1System} shows a schematic diagram of the proposed system architecture for solving the MAB problem using OAM. Here we illustrate the case where the number of arms is three, but the same principle applies in extending to a larger number of arms. 

Conventional laser sources generate beams that do not have orbital angular momentum. Technologically, methods to generate light with OAM from a plane wave or a Gaussian beam include the use of phase plates \cite{beijersbergen1994helical}, computer generated holograms (CGH) \cite{heckenberg1992generation}, or mode converters \cite{beijersbergen1993astigmatic, padgett1996experiment}. Spatial light modulators (SLMs) are widely utilized for this purpose, as they enable direct and tunable amplitude and/or phase modulation of an incoming light beam \cite{wang2012terabit}. The simplest and the most widely used method is a CGH-based approach implemented with an SLM and a 4f optical setup \cite{yao2011orbital}. In Fig. \ref{fig:1System}, a photon with a Gaussian spatial profile emitted from a laser is sent to a phase SLM, displaying a CGH pattern to generate OAM states, each carrying a phase factor $e^{i l\theta}$ which depends on the azimuthal angle $\theta$ and the OAM number $l$. $l$ could be any integer, but when all generated $l$ are expected to be positive, the output photon is described by the state: 
\begin{equation}
\begin{split}
SLM(\phi_1,\phi_2,\ldots,\phi_K)\ket{0} = \frac{1}{\sqrt{K}}\sum^{K}_{l=1}e^{i\phi_l}\ket{+l}
\end{split}
\end{equation}
where $\phi_1, \phi_2, \ldots, \phi_K$ depict phase changes associated with each OAM with $l$ values being +1, +2, $\ldots$, and $+K$, respectively, and $\ket{l}$ denotes the photon state with OAM value of $l$. That is to say, a single photon is emitted from the source system that contain $K$ OAM states with equal probability amplitude.

Meanwhile, a mirror causes flipping of the twisted structure of any given OAM; that is, the function of a beam splitter (BS) in the light propagation is represented by
\begin{equation} 
\begin{split} 
\ket{\Phi}\xrightarrow{1:1 \ beam \ splitter} \frac{1}{\sqrt{2}}\ket{\Phi}_{transmitted} + \frac{i}{\sqrt{2}}R\ket{\Phi}_{reflected},
\end{split}
\end{equation}
where $R$ represents flipping of OAM state, for example, $R\ket{+1}=\ket{-1}$. In the case $K=3$, we generate a photon state that carries equally $l=+1, +2, +3$ by setting $\phi_1=\phi_2=\phi_3=0$. That is, the output after SLM is given by $(1/\sqrt{3}) \times (\ket{+1}+\ket{+2}+\ket{+3})$. 

This photon is then transferred to an array of BSs and single photon detection system to examine which $l$-valued OAM is detected. Among a variety of methods in measuring the OAM of light \cite{leach2002measuring}, the system architecture shown in Fig. \ref{fig:1System} illustrates a method utilizing a hologram (HG) followed by a zeroth-order extraction system \cite{vaziri2003concentration}. In practical implementation, a zeroth-order extraction system could be free-space optics with spatial filtering or single-mode optical fibre. 

This hologram adds a phase factor of $e^{i l_{HG} \theta}$ to the state $\ket{l}$ with OAM $l$, which results in a transformation $\ket{l} \rightarrow \ket{l+l_{HG}}$. After injection into a zeroth-order extraction system, only an $l=0$ photon propagates in it. In other words, the zeroth-order extraction system acts as a filter to extract the $l=0$ component only. If the hologram induces a shift of OAM by $l_{HG}$ and a photon is detected by the subsequent photodetector, the OAM of the incoming photon is identified to be $l = -l_{HG}$. Based on this principle, in the system shown in Fig. \ref{fig:1System}, three holograms HG1, HG2, HG3 are arranged, which transform $\ket{l}$ into $\ket{l-1}, \ket{l-2}$, and $\ket{l-3}$, respectively. 

One remark here is that, although multiple BSs and holograms are employed in Fig. \ref{fig:1System}, more compact realization is indeed possible by, for example, a geometric optical transformation technique \cite{Lavery2012}, which has been extended to more than 50 OAM states \cite{Lavery2013}. The reason behind the introduction of the measurement architecture shown in Fig. \ref{fig:1System} regards the following procedure related to photon detections.

The output light is subjected to attenuators (ATT1, ATT2, ATT3) to control detection probabilities and a zeroth-order extraction system, followed by photodetectors (PD1, PD2, PD3). Based on the filtering by the zeroth-order extraction system, photon detection by PD1, PD2, and PD3 means observing OAM values of 1, 2, and 3, respectively. Photon detection by PD1 immediately means playing slot machine 1. Similarly, PD2 and PD3 are associated with the decision of playing slot machines 2 and 3, respectively. It should be emphasized that in this configuration, a machine is only selected if a photon is detected. 

Initially, since the probabilities of the detected photons to be measured by PD1, PD2, and PD3 are all equal to $1/3$, all machines are explored equally. Depending on the obtained results, the attenuation levels by ATT1, ATT2, ATT3 are updated.

After a single photon is detected by any photodetector, the selection yields eventual rewards from slot machines, and the results are registered into history $H(t)$. While referring to the history $H(t)$, the next decision is determined by following a certain policy of the player. The softmax policy is one of the most well-known feedback algorithms for the decision, which is also considered to accurately emulate the model of human decision making \cite{cohen2007should, daw2006cortical}. In the softmax policy, the player selects each machine based on a maximum likelihood estimation of the reward probability $\hat P_1(t),\hat P_2(t),\ldots,\hat P_K(t)$ and the probability of selecting machine $i$ is given by the following equation:
\begin{equation} 
\begin{split}
s_i(t+1) &= \frac{e^{\beta \hat P_i(t)}}{\displaystyle{\sum_{k=1}^{K}}e^{\beta \hat P_k(t)}}
\end{split}
\end{equation}
where $\beta$, which is also known as inverse temperature from analogy to statistical mechanics, is a parameter that influences the balance between exploration and exploitation. While optimal parameter $\beta$ depends on reward probabilities and some methods for tuning $\beta$ have been proposed \cite{cesa2017boltzmann}, this paper, for simplicity, set it to a constant value $\beta = 20$ based on a moderate tuning. The amplitude transmittance of attenuators (ATT1, ATT2, ATT3) are denoted by $d_1, d_2, d_3$, which are initially all one. These values are updated after every trial based on:
\begin{equation}
\begin{split}
d_i(t) = \sqrt{\frac{s_i(t)}{\displaystyle{\max_{k}}{s_k(t)}}}.
\end{split}
\end{equation}

In this way, $d_{i}(t)$ is revised as the time elapses so that the photon detection event is highly likely induced at the photodetector that corresponds to the best slot machine or the highest reward probability machine. For example, if slot machine 1 is the highest reward probability one, the transmittance of ATT1 should be higher while those of ATT2 and ATT3 should become smaller.

Here is a remark about the denominator of the right side of Eq. (5). The probability of detecting state $i$ is proportional to $d_i(t)^2$. Dividing each $d_i(t)^2$ by the same value $\displaystyle{\max_{k}}{s_k(t)}$ does not give any unintended bias to the detection probabilities, but transmission efficiency by the attenuators is kept high. That is, the loss of photons by the attenuators is minimized.

Finally, we discuss one more important remark regarding the architecture for solving the single-player, multi-armed bandit problem shown in Fig. 2. The principle maximizes the detection probability of the OAM state corresponding to the best machine. Actually, instead of reconfiguring the attenuators, we can accomplish the same functionality by reconfiguring the phase pattern displayed at the SLM located on the light source side. Indeed, this alternative way is directly and dynamically utilizing the high-dimensional property of OAM \cite{Pinnell2019}. This architecture, however, imposes a complex arbitration mechanism when we extend the principle to two-player situations in the following. That is, controlling the light source by a single player is indeed feasible, but the source management by two players is non-trivial. Instead, player-specific attenuator control does not impose any global server. For these reasons, we discuss the fundamental architecture shown in Fig. 2.

\subsubsection*{Simulation results for 1-player 3-armed bandit problem.}
\begin{figure}[ht] 
\centering
\includegraphics[width=\linewidth]{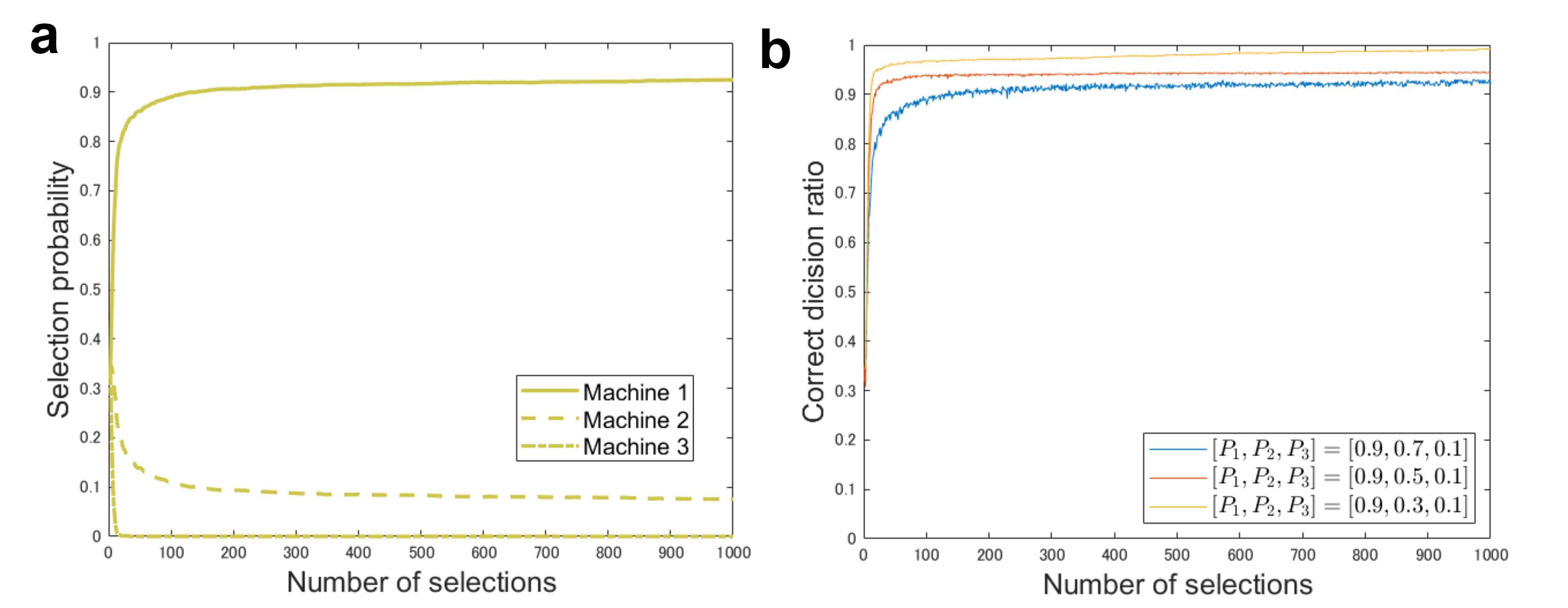}
\caption{Simulation results for solving 1-player 3-armed bandit problem with OAM system and softmax policy. \textbf{(a)} Probability of selecting each machine over 1000 times when $[P_1,P_2,P_3]=[0.9,0.7,0.1]$ averaged over 1000 repetitions. \textbf{(b)} Correct decision rate over 1000 repetitions in different reward environment.}
\label{fig:1Results}
\end{figure}

Figure \ref{fig:1Results} summarizes simulation results for the 1-player 3-armed bandit problem with the OAM system following the softmax policy. The solid, dashed, and dashed-dotted curves in Fig. \ref{fig:1Results}a show the time evolution of the selection probability of machine 1, 2, and 3, respectively, when the reward probability of slot machines are given by $[P_1, P_2, P_3] = [0.9, 0.7, 0.1]$. Here the number of repetitions is 1000. We can clearly observe that the probability of selecting the maximum reward probability machine, here machine 1, monotonically increases.

Figure \ref{fig:1Results}b examines the correct decision rate, which is referred to as CDR, defined by the number of selections of the highest reward probability machine over 1000 trials when the reward environment is configured differently. The blue, red, and yellow curves show the time evolution of CDR when the reward environment $[P_1, P_2, P_3]$ is given by $[0.9, 0.7, 0.1]$, $[0.9, 0.5, 0.1]$, and $[0.9, 0.3, 0.1]$, respectively. Here the maximum and minimum reward probabilities are commonly configured. As the difference between the maximum and the second maximum reward probability becomes smaller, the increase of CDR toward unity becomes slow. Nevertheless, we can observe that the monotonic increase of selecting the best machine in Figs. \ref{fig:1Results}a and \ref{fig:1Results}b. Since there is no theoretical limitation regarding the number of OAM states, the system configuration herein can be used for the probabilistic selection among a large number of selections. Note that the softmax policy itself is also scalable. 

\subsection*{Solving 2-player 3-armed bandit problem with OAM and quantum interference}

\subsubsection*{System architecture for solving 2-player 3-armed competitive bandit problem with OAM and quantum interference.}
\begin{figure}[ht] 
\centering
\includegraphics[width=\linewidth]{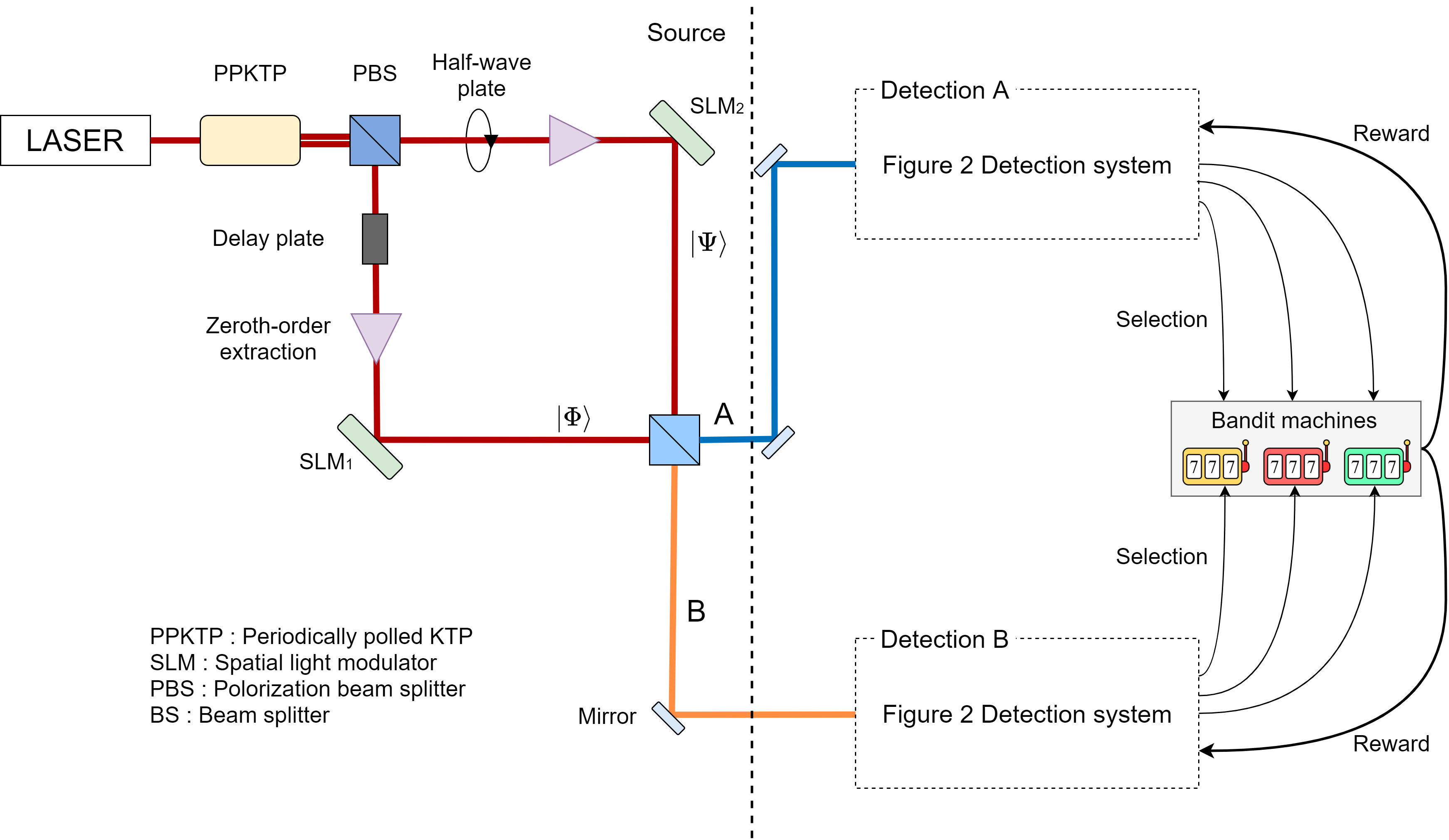}
\caption{System architecture for solving 2-player 3-armed bandit problem.}
\label{fig:2System}
\end{figure}

This section discusses stochastic selections of arms in the CMAB problem using photon pair OAM quantum states. The system presented in Fig. \ref{fig:1System} has been extended to the case of two players (Player A and B) by the architecture represented in Fig. \ref{fig:2System}. This time, the assumption is that the selection only happens when exactly one photon is detected simultaneously by each player on their photodetectors. 

In the source part, a photon pair is created by a nonlinear crystal such as a periodically poled KTP (PPKTP) and then subjected to an interferometer. One of the photon pair is supplied to the Detection A system, and the other goes to the Detection B system. The internal structure of Detection systems is the same as the one-player system depicted in Fig. 2. Thanks to the quantum interference, even though there is no explicit communication between the players, the detection results of the two photons are correlated with each other, as discussed in detail later.

In quantum research using light, it has been common to use quantum states based on properties such as polarization, spatial mode, and phase, but since the discovery of orbital angular momentum, many studies on quantum states using orbital angular momentum of light have been reported \cite{mair2001entanglement}. The availability of orbital angular momentum with an infinite number of states is very important in quantum research. In 2001, Mair \textit{et al.} used parametric down conversion (PDC) to study the generation of photon pairs in states with entangled orbital angular momentum \cite{mair2001entanglement}. Subsequently, a theoretical study of the change in orbital angular momentum during the PDC process was performed \cite{franke2002two}, and photon pairs with three entangled orbital angular momentum states were also studied \cite{vaziri2002experimental}. In the present study, we utilize quantum interference given by an extension of the Hong-Ou-Mandel effect \cite{hong1987measurement}.

\subsubsection*{Generation of OAM photon pair with quantum interference.}
\begin{figure}[ht] 
\centering
\includegraphics[width=\linewidth]{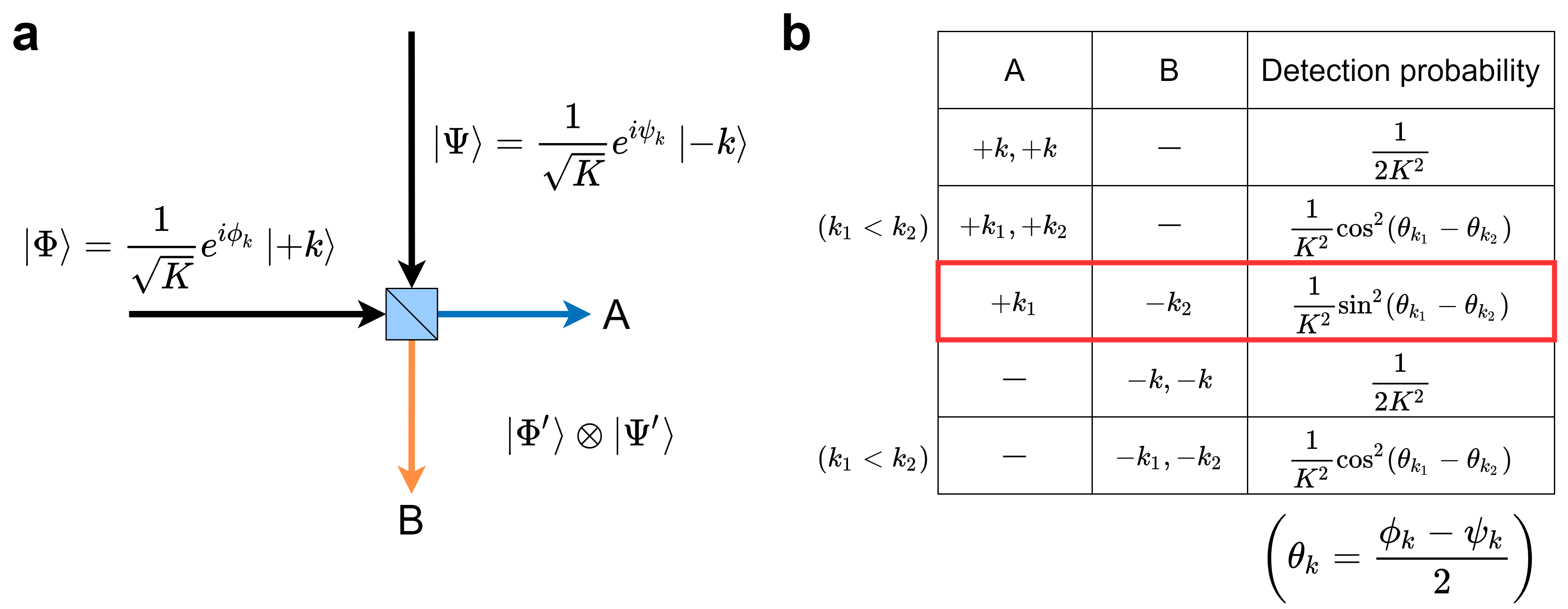}
\caption{\textbf{(a)} Photons with OAM states $\ket{\Phi},\ket{\Psi}$ go into BS as two inputs and go out to A and B. \textbf{(b)} Probability of detecting certain OAM states at each side of a beam splitter A and B. Some of the probabilities are constant, while the others depend on the phase difference between two input states. The probability of a pair of photons being detected at different sides is marked by the red frame.}
\label{fig:HOM}
\end{figure}
\begin{figure}[ht] 
\centering
\includegraphics[width=16.5cm]{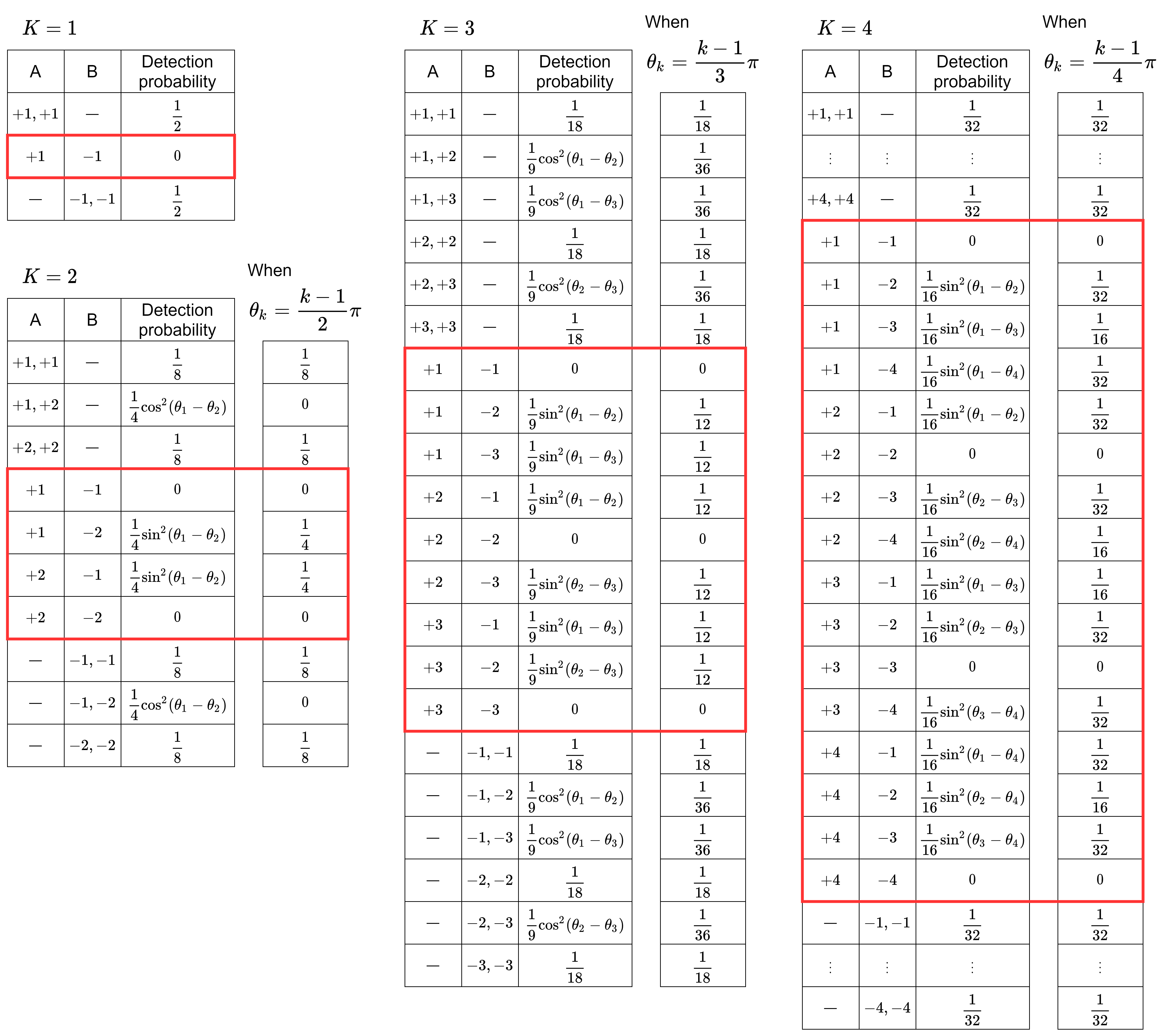}
\caption{Probability of detecting certain OAM states at either sides of the beam splitter with $K$ ranging from 1 to 4. The probability of a pair of photons being detected at different sides are marked by the red frame.}
\label{fig:HOM1234}
\end{figure}

Hong-Ou-Mandel effect has been well studied for two identical photons always detected together in the same output path when they enter into a 1:1 beam splitter \cite{hong1987measurement}. We extend the description of this phenomenon for multiple-OAM states carrying input photons. When OAM states of input photon $\ket{\Phi}$ is sent to the beam splitter, transmitted term A and reflected term B can be described with the following forms:
\begin{equation}
\begin{split}
\ket{\Phi}\xrightarrow{1:1 \ beam \ splitter} \frac{1}{\sqrt{2}}\ket{\Phi}_A + \frac{i}{\sqrt{2}}R\ket{\Phi}_B.
\end{split}
\end{equation}
where $R$ represents flipping of OAM state, for example, $R\ket{+1}=\ket{-1}$. As shown in Fig. \ref{fig:HOM}a, when OAM states of input photons are $\ket{\Phi},\ket{\Psi}$ on the two BS inputs, the output state $\ket{\Phi'}\otimes\ket{\Psi'}$ can be described with the following forms:
\begin{equation}
\begin{split}
\ket{\Phi'}\otimes\ket{\Psi'} 
&= \left(\frac{1}{\sqrt{2}}\ket{\Phi}_A + \frac{i}{\sqrt{2}}R\ket{\Phi}_B \right) \otimes \left(\frac{i}{\sqrt{2}}R\ket{\Psi}_A + \frac{1}{\sqrt{2}}\ket{\Psi}_B \right) \\
&= \left(\frac{i}{2}\ket{\Phi}_A\otimes R\ket{\Psi}_A\right) + \left(\frac{1}{2}\ket{\Phi}_A\otimes\ket{\Psi}_B - \frac{1}{2}R\ket{\Psi}_A\otimes R\ket{\Phi}_B\right) + \left(\frac{i}{2}R\ket{\Phi}_B\otimes\ket{\Psi}_B\right).
\end{split}
\label{math:bs1}
\end{equation}
With $K$ being the number of OAM used in the system, the input states $\ket{\Phi},\ket{\Psi}$ can be set to
\begin{equation}
\begin{split}
\ket{\Phi} = \frac{1}{\sqrt{K}}\sum^{K}_{k=1}e^{i\phi_k}\ket{+k}, \ \ \ \ket{\Psi} = \frac{1}{\sqrt{K}}\sum^{K}_{k=1}e^{i\psi_k}\ket{-k},
\end{split}
\end{equation}
considering that the two photons have the same polarization, wavelength, and are synchronized on the beam splitter. Each term of the output state given by Eq. \ref{math:bs1} is described by the following:
\begin{equation} 
\begin{split}
\ket{\Phi}_A\otimes R\ket{\Psi}_A 
&=\left( \sum_{k=1}^{K} \frac{1}{\sqrt{K}}e^{i\phi_k} \ket{+k}_A\right) \otimes \left(\sum_{k=1}^{K} \frac{1}{\sqrt{K}} e^{i\psi_k} \ket{+k}_A\right)\\
&= \sum_{k=1}^{K}\frac{1}{K}e^{i(\phi_k+\psi_k)}\ket{+k}_A\otimes\ket{+k}_A + \sum_{k_1 < k_2}^{K}\frac{1}{K}\left(e^{i(\phi_{k_1}+\psi_{k_2})}+e^{i(\psi_{k_1}+\phi_{k_2})}\right)\ket{+k_1}_A\otimes\ket{+k_2}_A
\end{split}
\end{equation}
\begin{equation}
\begin{split}
\ket{\Phi}_A\otimes\ket{\Psi}_B - R\ket{\Psi}_A\otimes R\ket{\Phi}_B
&= \sum_{k=1}^{K} \frac{1}{\sqrt{K}}e^{i\phi_k} \ket{+k}_A \otimes \sum_{k=1}^{K} \frac{1}{\sqrt{K}} e^{i\psi_k} \ket{-k}_B - \sum_{k=1}^{K} \frac{1}{\sqrt{K}}e^{i\psi_k} \ket{+k}_A \otimes \sum_{k=1}^{K} \frac{1}{\sqrt{K}} e^{i\phi_k} \ket{-k}_B\\
&= \sum_{k_1=1}^{K}\sum_{k_2=1}^{K}\frac{1}{K}e^{i(\phi_{k_1}+\psi_{k_2})}\ket{+k_1}_A\otimes\ket{-k_2}_B - \sum_{k_1=1}^{K}\sum_{k_2=1}^{K}\frac{1}{K}e^{i(\psi_{k_1}+\phi_{k_2})}\ket{+k_1}_A\otimes\ket{-k_2}_B \\
&= \sum_{k_1=1}^{K}\sum_{k_2=1}^{K}\frac{1}{K}\left(e^{i(\phi_{k_1}+\psi_{k_2})}-e^{i(\psi_{k_1}+\phi_{k_2})}\right)\ket{+k_1}_A\otimes\ket{-k_2}_B.
\end{split}
\end{equation}
Therefore, the output state $\ket{\Phi'}\otimes\ket{\Psi'}$ is given by the following terms:
\begin{equation}
\begin{split}
\ket{\Phi'}\otimes\ket{\Psi'} &= 
\sum_{k=1}^{K}\frac{i}{2K}e^{i(\phi_k+\psi_k)}\ket{+k}_A\otimes\ket{+k}_A \\
&+\sum_{k_1 < k_2}^{K}\frac{i}{2K}\left(e^{i(\phi_{k_1}+\psi_{k_2})}+e^{i(\psi_{k_1}+\phi_{k_2})}\right)\ket{+k_1}_A\otimes\ket{+k_2}_A \\
&+\sum_{k_1=1}^{K}\sum_{k_2=1}^{K}\frac{1}{2K}\left(e^{i(\phi_{k_1}+\psi_{k_2})}-e^{i(\psi_{k_1}+\phi_{k_2})}\right)\ket{+k_1}_A\otimes\ket{-k_2}_B \\
&+\sum_{k=1}^{K}\frac{i}{2K}e^{i(\phi_k+\psi_k)}\ket{-k}_B\otimes\ket{-k}_B \\
&+\sum_{k_1 < k_2}^{K}\frac{i}{2K}\left(e^{i(\phi_{k_1}+\psi_{k_2})}+e^{i(\psi_{k_1}+\phi_{k_2})}\right)\ket{-k_1}_B\otimes\ket{-k_2}_B.
\end{split}
\end{equation}
Correspondingly, the probability of detecting the same state at the same side, that is $\ket{+k}_A\otimes\ket{+k}_A$ or $\ket{-k}_B\otimes\ket{-k}_B$, is given by
\begin{equation}
\begin{split}
2\cdot\left|\frac{i}{2K}e^{i(\phi_k+\psi_k)}\right|^2 = \frac{1}{2K^2}.
\end{split}
\end{equation}
By introducing parameters $\theta_k=\frac{\phi_k-\psi_k}{2}$, which depends on the phase difference of two input states, the probability of detecting different states on the same side, that is $\ket{+k_1}_A\otimes\ket{+k_2}_A$ or $\ket{-k_1}_B\otimes\ket{-k_2}_B$, is given by
\begin{equation}
\begin{split}
\left|\frac{i}{2K}\left(e^{i(\phi_{k_1}+\psi_{k_2})}+e^{i(\psi_{k_1}+\phi_{k_2})}\right)\right|^2 = \frac{1}{K^2}\cos^2(\theta_{k_1}-\theta_{k_2}),
\end{split}
\end{equation}
and finally the probability of detecting pair of states on different sides, that is $\ket{+k_1}_A\otimes\ket{-k_2}_B$, is given by
\begin{equation}
\begin{split}
\left|\frac{1}{2K}\left(e^{i(\phi_{k_1}+\psi_{k_2})}-e^{i(\psi_{k_1}+\phi_{k_2})}\right)\right|^2 = \frac{1}{K^2}\sin^2(\theta_{k_1}-\theta_{k_2}).
\end{split}
\end{equation}
Figure \ref{fig:HOM}b summarizes the probability of detecting each output state, while Fig. \ref{fig:HOM1234} shows all the probabilities with $K$ ranging from 1 to 4. The probabilities depend only on $\theta_k$, which can be tuned by controlling the SLM phases $\phi_k$ and $\psi_k$. 

A pair of photons being detected on both sides is displayed with the red frames in Fig. \ref{fig:HOM1234}, which are utilized as selections by the two players. What is remarkable is that the probability of detecting the same states at different sides is always zero because the probability term $\sin^2(\theta_k-\theta_k)$ is always equal to zero. For $K=1$, this phenomenon corresponds to what is known as the Hong-Ou-Mandel effect. As the detected OAM states correspond to the selection of players, the probability of both players selecting the same machine is only limited by experimental constraints such as multiple pair generation, meaning that conflict-free decisions are accomplished.

The probabilities described in the red frames include the probabilities of detecting different states by the two players. It is remarkable that these probabilities can take equal value when $K$ is less than or equal to three. For example, when $K = 2$, by assigning $\theta_1 = 0$ and $\theta_2 = \pi/2$, all such probabilities becomes $1/4$. Similarly, when $K = 3$, by setting $(\theta_1, \theta_2, \theta_3) = (0, \pi/3, 2\pi/3)$, the probabilities are all $1/12$. Namely, all arm combinations except selecting the same arm are selected equally. Note, however, that when $K$ is larger or equal to four, we cannot perfectly equalize these probabilities by only tuning $\theta_1,\theta_2,\ldots,\theta_K$. This point is discussed in the \textit{Discussion} section. In this study, we focus on the case when $K = 3$ because the equivalent selection of pairs is ensured, as discussed above.

\subsubsection*{Simulation results for 2-player 3-armed bandit problem.}
\begin{figure}[ht] 
\centering
\includegraphics[width=\linewidth]{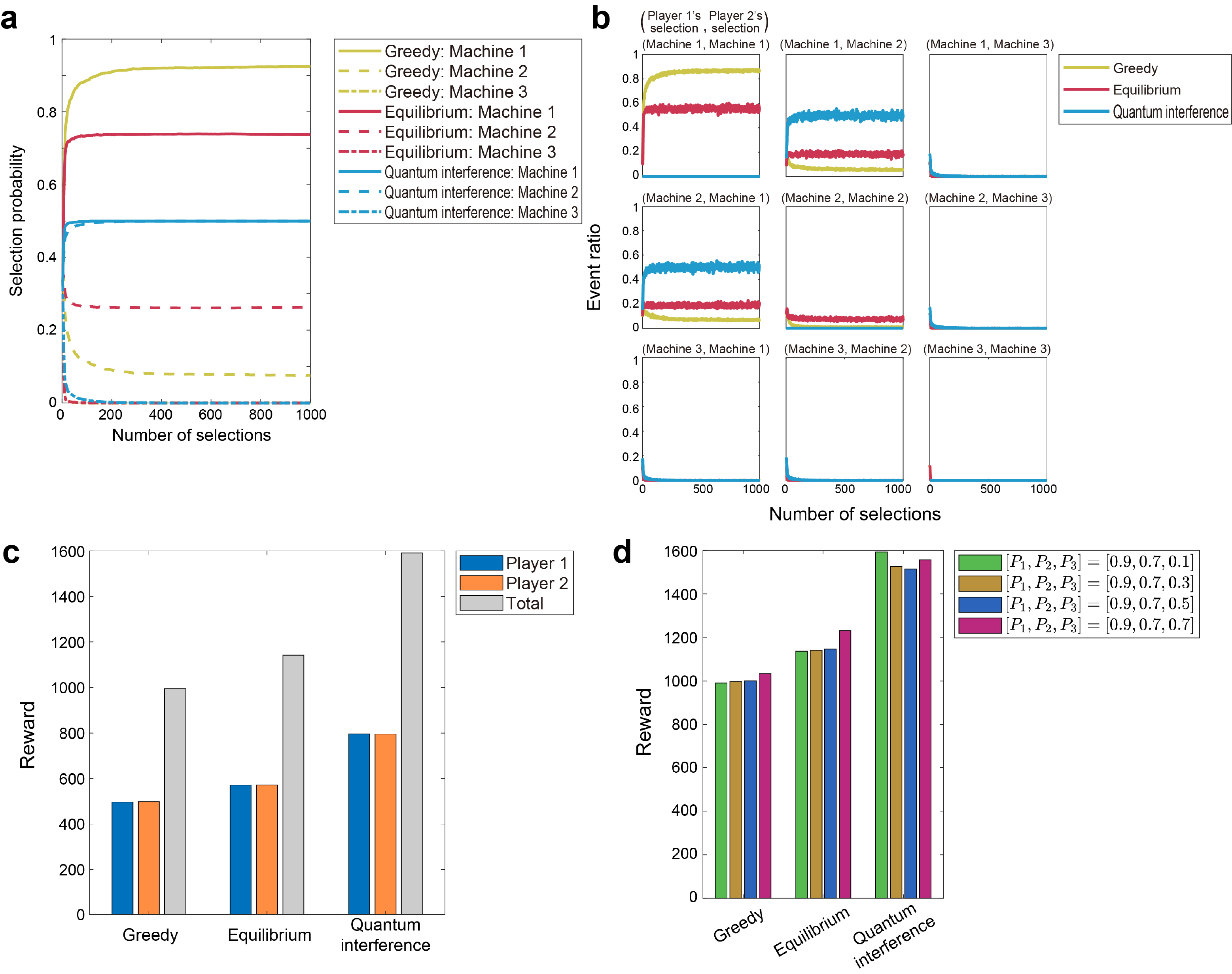}
\caption{Simulation results for solving 2-player 3-armed competitive bandit problem with greedy, equilibrium and quantum interference policies. \textbf{(a)} Probability of selecting each machine over 1000 selections when $[P_1,P_2,P_3]=[0.9,0.7,0.1]$ averaged over 1000 repetitions. \textbf{(b)} Ratio of each selection pair over 1000 selections and 1000 repetitions. \textbf{(c)} Average rewards of player 1, player 2, and total rewards over 1000 selections with $[P_1,P_2,P_3]=[0.9,0.7,0.1]$ averaged over 1000 repetitions. \textbf{(d)} Average total rewards with different reward probability over 1000 selections and 5000 repetitions.}
\label{fig:2Results}
\end{figure}
In the CMAB problem in the present study, the rewards are equally split among the players who selected the same machine; that is, the decision conflict by multiple players reduces the individual benefit. Furthermore, total rewards are reduced because of the conflicted choice. 

Here we begin with a brief overview of the two-player decision-making situations by a game-theoretic formalism \cite{nash1950equilibrium} while mentioning its intuitive implications. We denote $P_{k^{*}}, P_{k^{**}}, P_{k^{***}}$ respectively the first, second, and third highest reward probability. First, when $P_{k^{*}} > 2\times P_{k^{**}}$, the situation of both players selecting machine 1 is the only Nash equilibrium. That is,  conflict is unavoidable if both players act in a greedy manner because the best machine is far better than the other machines.

Second, when $P_{k^{*}}<2 \times P_{k^{**}}$, Nash equilibrium is  achieved when player 1 chooses the best machine (machine $k^{*}$), and player 2 selects the second-best machine (machine $k^{**}$), and vice versa. That is, conflicting decisions are avoided because changing the player’s decision decreases his/her reward. However, there is a problem from the viewpoint of equality, as one of the players can keep selecting the higher reward machines while the other is locked with the lower reward decisions.

Third, there exists another symmetric Nash equilibrium with a mixed strategy, meaning that they select each machine with a certain probability. The details are described in the \textit{Methods} section. Intuitively speaking, by this mixed strategy, both players sometimes intentionally refrain from choosing the best machine. Therefore, sometimes, decision conflicts can be avoided. Indeed, Lai \textit{et al.} successfully utilized a mixed strategy in dynamic channel selection in communication systems \cite{lai2010cognitive}. However, it should be remarked that perfect conflict avoidance cannot be ensured by mixed strategies.

In order to quantitatively evaluate the performance differences among different policies, we compare the quantum interference system with the following two policies. One is a greedy policy where both players take greedy actions as if they are playing alone. The second is an equilibrium policy where both players try to achieve the symmetric Nash equilibrium by a mixed strategy. The details are described in the \textit{Methods} section.

Figure \ref{fig:2Results} shows the results for solving the 2-player 3-armed bandit problem. Figure \ref{fig:2Results}a shows how the selection probabilities of both players evolve with each policy. With the greedy policy, reminding that machine 1 has the highest reward probability of 0.9, its selection probability approaches almost 1 for both players, as in the case of a single player. For the equilibrium policy, the selection probabilities of the two most rewarding machines 1 and 2 converge to the probabilities defined by the mixed strategy. With the quantum interference strategy, however, machine 1 and machine 2 are selected with equal probability by both players. Figure \ref{fig:2Results}b shows the ratio of each selection combination from both players. The greedy policy is associated with a large number of conflicts as both players almost only select machine 1, while the equilibrium policy reduces the number of conflicts to some extent as the selections are distributed. Finally, the quantum interference policy completely avoids conflicts. The final rewards with such selections are shown in Fig. \ref{fig:2Results}c, for each player and for the total attributed reward. We observe that the quantum interference policy achieves almost ideal total rewards as well as equality between players. By contrast, the total reward by the greedy and the equilibrium policies becomes small compared with the quantum interference policy because they suffer from unavoidable decision conflicts. 

Figure \ref{fig:2Results}d shows how the final reward of each policy varies when the reward probabilities of the three machines are modified. In greedy and equilibrium policies, the total reward changes due to the rate of selection of the lowest rewarding machine 3 in the exploration phase. On the other hand, with the quantum interference policy, the larger the difference between the reward probabilities of machine 2 and machine 3, the easier it is to determine the top two machines, and the higher the final total reward; despite this, the difference in total reward is mild in comparison with the difference with the other two policies.

\section*{Discussion}
In this study, we show that we can benefit from the high dimensionality of OAM for scalability in solving multi-armed bandit problems. Furthermore, appropriate quantum interference constructions lead to achieving high rewards while maintaining a fair repartition between two players in competitive bandit problem situations. The total reward optimization is guaranteed by the selections of the two best machines by the two players in a non-conflicting manner, while the fair repartition is guaranteed by the equal probabilities of selection among players through quantum interference.

The main assumption is the simultaneous detection of exactly one photon for all players. In the proposed optical design, this is for the purpose of the extended Hong-Ou-Mandel effect or quantum interference that guarantees that identical photons go to the same side of the beam splitter, at the price of a post-selection of half of all photon pairs. While this is a strong constraint for potential applications, this design is only an example, and nothing forbids the obtention of the target state with other designs that do not rely on post-selection. 

Regarding the extension to more arms, the current design is limited to three arms due to fundamental constraints (lack of enough degrees of freedom to constrain the 2-photon state). This may be solved by allowing to tune the relative amplitude between each OAM with the SLM and/or additional mechanisms. Once again, the goal of the setup presented in this study is only to present the principle of utilizing OAM for multiple arms in MAB and quantum interference for competitive decision-making. We believe that the extension to many arms is a technological problem without theoretical constraints \cite{Lavery2013}.

The next discussion point is about security. The two-player CMAB system herein intends to let the players directly influence the detection probability via the attenuation amplitude in front of the detectors. While this architecture ensures independent machine selection and revision of the attenuation among the players, it presents one fundamental weakness: if a player only wants to select the highest rewarding arm, then the attenuation will be maximized for the lower arms, only letting photons reach the corresponding detector. However, this situation is easily identifiable by the other player, who can recognize that the probability of selecting a particular machine decreases. The solution for that player is straightforward: attenuate more the second-best arm too to correct the imbalance (in case of slight inequality), which is equivalent to not playing anymore if the other player completely blocks the other photons. 

This brings the following discussion point about the photon utilization efficiency. In this study, only simultaneous detection of exactly one photon for all detectors of each and every player triggers the selection of arms from both players. The reason is to implement the post-selection of output states where one photon goes for both players instead of two photons for only one player. With the current operation principle based on quantum interference summarized in Fig. 6, half of all photon pairs are strictly unusable for the players. Although such a loss is unavoidable, further photon losses are induced in the system architecture shown in Fig. 4 because of the multiple BSs. As discussed earlier, this part can be improved by technological methods developed in the literature \cite{Lavery2012},\cite{Lavery2013}. 

\section*{Conclusion}
To overcome the scalability limitations in the former single-photon-based decision making that relies on two orthogonal polarizations to resolve the two-armed bandit problem, we associate orbital angular momentum of photons to individual arms, which theoretically allows ideal scalability. When multiple players are involved, conflict of decisions becomes a serious issue, which is known as the competitive multi-armed bandit problem. Formerly, polarization-entangled photons have been shown to realize conflict-free decision making in two-player, two-armed situations; however, its arm-scalability is limited to only two. In this study, by extending the Hong-Ou-Mandel effect to more than two states, we theoretically establish an experimental configuration able to generate quantum interference among states with orbital angular momentum and conditions that provide conflict-free selections. We numerically examine total rewards regarding two-player, three-armed bandit problems, for which the proposed principle accomplishes almost the theoretical maximum, which is greater than a conventional mixed strategy intending to realize Nash equilibrium. This study paves a way toward photon-based intelligent systems as well as extending the utility of the high dimensionality of orbital angular momentum of photons and quantum interference in artificial intelligence domains.

\section*{Methods}
\subsection*{Detail algorithm of greedy policy, equilibrium policy, and entanglement policy.}
\subsubsection*{Greedy policy (Strategty for single player MAB).}
Both players independently decide the probability of selecting each machine at each round. The algorithm is based on the softmax policy \cite{10.5555/551283} and the probability of selecting machine $i$ at round $t$ is given by the following equation:
\begin{equation}
\begin{split}
s_k(t+1) &= \frac{e^{\beta \hat P_k(t)}}{\displaystyle{\sum_{n=1}^{K}}e^{\beta \hat P_n(t)}}.
\end{split}
\end{equation}
If there exists a non-selected machine, all the machines are selected randomly with the same probability. Otherwise, $\hat P_k(t)=\frac{w_k(t)}{w_k(t)+l_k(t)}$ when the player has been rewarded $w_k(t)$ times and not rewarded $l_k(t)$ times from machine $k$. As a moderate tuning, $\beta$ is set to be 20 in this study.

\subsubsection*{Equilibrium policy.}
\begin{table}[ht]
\centering
\begin{tabular}{|c|c|c|c|c|}
\hline
\multicolumn{2}{|l|}{\multirow{2}{*}{\begin{tabular}[c]{@{}l@{}}Expected reward for \\ (Player 1, Player 2)\end{tabular}}} & \multicolumn{3}{l|}{Selection of Player 2} \\ \cline{3-5} 
\multicolumn{2}{|l|}{} & Machine 1 & Machine 2 & Machine 3 \\ \hline
\multirow{3}{*}{Selection of Player 1} & Machine 1 & $(P_1/2,P_1/2)$ & $(P_1,P_2)$ & $(P_1,P_3)$ \\
\cline{2-5} & Machine 2 & $(P_2,P_1)$ & $(P_2/2,P_2/2)$ & $(P_2,P_3)$ \\ 
\cline{2-5} & Machine 3 & $(P_3,P_1)$ & $(P_3,P_2)$ & $(P_3/2,P_3/2)$ \\ \hline
\end{tabular}
\caption{Profit table of expected reward. Reward is splitted when the selection is in conflict between two players.}
\label{table:ProfitTable}
\end{table}
\noindent Table\ref{table:ProfitTable} represents the profit table of the expected reward in a single selection. In Nash equilibrium, no player has anything to gain by changing only their own strategy. In Nash equilibrium, the strategy may be selecting one particular machine, but it could also be a strategy such that multiple machines are probabilistically chosen. In the situation of Table \ref{table:ProfitTable}, strategies can be defined with the probabilities of selecting machine 1, machine 2, machine 3, or $\alpha_1,\alpha_2,\alpha_3$ for player A and $\beta_1,\beta_2,\beta_3$ for player B. In what follows, the notations $k^{*},k^{**},k^{***}$ represent indices of the first, the second, and the third best machine, respectively. Nash equilibriums are summarized as shown below:
\begin{itemize}
\item in case $P_{k^{*}}>2P_{k^{**}}$
\begin{itemize}
\item[$\diamond$] $(\alpha_{k^{*}},\alpha_{k^{**}},\alpha_{k^{***}}) = (\beta_{k^{*}},\beta_{k^{**}},\beta_{k^{***}}) = (1,0,0)$
\end{itemize}
\item in case $P_{k^{*}}<2P_{k^{**}}$ and $P_{k^{*}}P_{k^{**}}/Q>\frac{2}{5}$
\begin{itemize}
\item $(\alpha_{k^{*}},\alpha_{k^{**}},\alpha_{k^{***}}) = (1,0,0), \ (\beta_{k^{*}},\beta_{k^{**}},\beta_{k^{***}}) = (0,1,0)$
\item $(\alpha_{k^{*}},\alpha_{k^{**}},\alpha_{k^{***}}) = (0,1,0), \ (\beta_{k^{*}},\beta_{k^{**}},\beta_{k^{***}}) = (1,0,0)$
\item[$\diamond$] $(\alpha_{k^{*}},\alpha_{k^{**}},\alpha_{k^{***}}) = (\beta_{k^{*}},\beta_{k^{**}},\beta_{k^{***}}) = (\frac{2P_{k^{*}}-P_{k^{**}}}{P_{k^{*}}+P_{k^{**}}}, \frac{2P_{k^{**}}-P_{k^{*}}}{P_{k^{*}}+P_{k^{**}}},0)$
\end{itemize}
\item in case $P_{k^{*}}<2P_{k^{**}}$ and $P_{k^{*}}P_{k^{**}}/Q<\frac{2}{5}$
\begin{itemize}
\item $(\alpha_{k^{*}},\alpha_{k^{**}},\alpha_{k^{***}}) = (1,0,0), \ (\beta_{k^{*}},\beta_{k^{**}},\beta_{k^{***}}) = (0,1,0)$
\item $(\alpha_{k^{*}},\alpha_{k^{**}},\alpha_{k^{***}}) = (0,1,0), \ (\beta_{k^{*}},\beta_{k^{**}},\beta_{k^{***}}) = (1,0,0)$
\item[$\diamond$] $(\alpha_{k^{*}},\alpha_{k^{**}},\alpha_{k^{***}}) = (\beta_{k^{*}},\beta_{k^{**}},\beta_{k^{***}}) = (2 - \frac{5P_{k^{**}}P_{k^{***}}}{Q}, 2 - \frac{5P_{k^{***}}P_{k^{*}}}{Q}, 2 - \frac{5P_{k^{*}}P_{k^{**}}}{Q})$
\end{itemize}
\end{itemize}
where $Q = P_{k^{*}}P_{k^{**}}+P_{k^{**}}P_{k^{***}}+P_{k^{***}}P_{k^{*}}$. With the equilibrium policy, both players try to achieve symmetric Nash equilibrium, which is represented with the shape $\diamond$ above, under the situation that reward probabilities are not quite sure. In the simulation algorithm, each player decides which machines are better and which Nash equilibrium to achieve based on their own maximum likelihood estimation of reward probabilities. In an actual algorithm, the parameters of player 1 are calculated as below with $\hat k^{*}, \hat k^{**}, \hat k^{***}$ respectively representing machine indices with the first, the second, and the third highest estimated reward probability:
\begin{itemize}
\item in case $\hat P_{\hat k^{*}}>2\hat P_{\hat k^{**}}$
\begin{itemize}
\item $(\alpha^{*},\alpha^{**},\alpha^{***}) = (1,0,0)$
\end{itemize}
\item in case $\hat P_{\hat k^{*}}>2\hat P_{\hat k^{**}}$ and $\hat P_{\hat k^{*}}\hat P_{\hat k^{**}}/Q>\frac{2}{5}$
\begin{itemize}
\item $(\alpha^{*},\alpha^{**},\alpha^{***}) = (\frac{2\hat P_{\hat k^{*}}-\hat P_{\hat k^{**}}}{\hat P_{\hat k^{*}}+\hat P_{\hat k^{**}}}, \frac{2\hat P_{\hat k^{**}}-\hat P_{\hat k^{*}}}{\hat P_{\hat k^{*}}+\hat P_{\hat k^{**}}},0)$
\end{itemize}
\item in case $\hat P_{\hat k^{*}}<2\hat P_{\hat k^{**}}$ and $\hat P_{\hat k^{*}}\hat P_{\hat k^{**}}/Q<\frac{2}{5}$
\begin{itemize}
\item $(\alpha^{*},\alpha^{**},\alpha^{***}) = (2 - \frac{5\hat P_{\hat k^{**}}\hat P_{\hat k^{***}}}{Q}, 2 - \frac{5\hat P_{\hat k^{***}}\hat P_{\hat k^{*}}}{Q}, 2 - \frac{5\hat P_{\hat k^{*}}\hat P_{\hat k^{**}}}{Q})$
\end{itemize}
\end{itemize}
where $Q = \hat P_{\hat k^{*}}\hat P_{\hat k^{**}}+\hat P_{\hat k^{**}}\hat P_{\hat k^{***}}+\hat P_{\hat k^{***}}\hat P_{\hat k^{*}}$, and $\hat P_{\hat k^{*}},\hat P_{\hat k^{**}},\hat P_{\hat k^{***}}$ represent the first, the second, and the third highest estimated reward probability. The parameters of player 2 are also calculated in the same way with the different reward probability estimations. The probability of selecting each machine is calculated as below:
\begin{equation}
\begin{split}
s_{\hat k^{*}}(t+1) &= \alpha^{*}\pi(P_{\hat k^{*}}=P_{k^{*}}|H(t)) + \alpha^{**}\pi(P_{\hat k^{*}}=P_{k^{**}}|H(t)) + \alpha^{***}\pi(P_{\hat k^{*}}=P_{k^{***}}|H(t)) \\
s_{\hat k^{**}}(t+1) &= \alpha^{*}\pi(P_{\hat k^{**}}=P_{k^{*}}|H(t)) + \alpha^{**}\pi(P_{\hat k^{**}}=P_{k^{**}}|H(t)) + \alpha^{***}\pi(P_{\hat k^{**}}=P_{k^{***}}|H(t)) \\
s_{\hat k^{***}}(t+1) &= \alpha^{*}\pi(P_{\hat k^{***}}=P_{k^{*}}|H(t)) + \alpha^{**}\pi(P_{\hat k^{***}}=P_k{^{**}}|H(t)) + \alpha^{***}\pi(P_{\hat k^{***}}=P_{k^{***}}|H(t))
\end{split}
\end{equation}
where, $\pi(P_{\hat k^{*}}=P_{k^{*}}|H(t))$ represents the probability of machine $\hat k^{*}$ to have the highest reward probability from the estimation based on the softmax policy. Here, $\pi(P_a>P_b>P_c|H(t))$ represents the probability of reward probabilities being $P_a>P_b>P_c$ under this estimation and it is calculated as below:
\begin{equation}
\begin{split}
\pi(P_a>P_b>P_c|H(t)) = \frac{e^{\beta \hat P_a}}{e^{\beta \hat P_a}+e^{\beta \hat P_b}+e^{\beta \hat P_c}}\cdot \frac{e^{\beta \hat P_b}}{e^{\beta \hat P_b}+e^{\beta \hat P_c}}.
\end{split}
\end{equation}
Therefore, probabilities of $P_a$ being the first, the second, and the third best machine under estimation with softmax policy are:
\begin{equation}
\begin{split}
\pi(P_a=P_{k^{*}}|H(t)) = \pi(P_a>P_b>P_c|H(t)) + \pi(P_a>P_c>P_b|H(t)) \\
\pi(P_a=P_{k^{**}}|H(t)) = \pi(P_b>P_a>P_c|H(t)) + \pi(P_c>P_a>P_b|H(t)) \\
\pi(P_a=P_{k^{***}}|H(t)) = \pi(P_b>P_c>P_a|H(t)) + \pi(P_c>P_b>P_a|H(t)).
\end{split}
\end{equation}

\subsubsection*{Quantum interference policy.}
In the quantum interference policy, both players try to select both the first and the second-best machine with the same probability to achieve fairness between the two players. Therefore, the probabilities of selecting machines are given by the following equation:
\begin{equation}
\begin{split}
s_{k}(t+1) &= \frac{1}{2}\left(\pi(P_k=P_{k^{*}}|H(t)) + \pi(P_k=P_{k^{**}}|H(t))\right).
\end{split}
\end{equation}

\bibliography{main}



\section*{Acknowledgements}
This work was supported in part by the CREST Project (JPMJCR17N2) funded by the Japan Science and Technology Agency and Grants-in-Aid for Scientific Research (JP20H00233) funded by the Japan
Society for the Promotion of Science.

\section*{Author contributions statement}
M.N., N.C., and G.B. directed the project. T.A., N.C., G.B., S.H., and M.N designed the system architecture. T.A. and N.C conducted physical modeling and numerical performance evaluations. N.C., G.B., S.H., and R.H. examined technological constraints. All authors discussed the results. T.A., N.C., and M.N. wrote the manuscript. All authors reviewed the manuscript.


\section*{Additional information}
Correspondence and requests for materials should be addressed to T.A. and M.N. \\
\textbf{Competing interests}: The authors declare no competing interests.



\end{document}